\documentclass{aastex63}
\usepackage{placeins, verbatim}

\newcommand{\num}[1]{\textcolor{black}{#1}} 

\begin{document}

\title{Microlensing Constraints on the Stellar and Planetary Mass Functions}
\author{Jennifer C. Yee}
\affiliation{Center for Astrophysics $|$ Harvard \& Smithsonian, 60 Garden St.,Cambridge, MA 02138, USA}
\email{jyee@cfa.harvard.edu}
\author{Scott J. Kenyon}
\affiliation{Center for Astrophysics $|$ Harvard \& Smithsonian, 60 Garden St.,Cambridge, MA 02138, USA}

\begin{abstract}
The mass function (MF) of isolated objects measured by microlensing consists of both a stellar and a planetary component. We compare the microlensing MFs of \citet{Gould22_FFPs} and \citet{Sumi23_FFPs} to other measurements of the MF. The abundance of brown dwarfs in the \citet{Sumi23_FFPs} stellar MF is consistent with measurements from the local solar neighborhood \citep{Kirkpatrick2024}. Microlensing free-floating planets  ($\mu$FFPs) may may be free-floating or orbit host stars with semimajor axes $a\gtrsim 10~\mathrm{au}$ and therefore can constrain the populations of both free-floating planetary-mass objects and wide-orbit planets. Comparisons to radial velocity and direct imaging planet populations suggest that either most of the $\mu$FFP population with masses $>1~M_{\rm Jup}$ is bound to hosts more massive than M dwarfs or some fraction of the observed bound population actually comes from the low-mass tail of the stellar population. The $\mu$FFP population also places strong constraints on planets inferred from debris disks and gaps in protoplanetary disks observed by ALMA.

\end{abstract}

\section{Introduction}

Measurements of the microlensing free-floating planet ($\mu$FFP) population \citep{Mroz20FFP,Ryu21_kb2820,Gould22_FFPs,Koshimoto23_FFPs,Sumi23_FFPs} have broad implications for both the stellar and planetary mass functions (MFs) and theories of their formation. Current theories for the formation of planets and stars differ in significant ways. Stars grow at the center of a collapsing molecular cloud; planets grow within a disk of material orbiting a newly-formed star \citep[e.g.][]{Shu1987,Drazkowska2023,Pineda2023}. These systems evolve as planets migrate, interact, or get ejected from their parent systems either during or after the planet formation process. We consider the connections between the $\mu$FFP population, the wide-orbit planet population, and the stellar initial MF (IMF), which have implications for constraining the evolution of the planetary systems and theories of their formation.

The properties of the $\mu$FFP population are inferred by measuring deviations from extrapolating the stellar MF to planetary masses. Because the two populations are measured together, we can decompose objects into separate populations of stars and planets
and then investigate the mass range where the stellar and planetary mass functions converge. 
Unlike most microlensing planets, $\mu$FFPs do not have a microlensing signal from a host star. They could be true free-floating objects or planets on very wide orbits with semimajor axes $\gtrsim$ 10~au. Despite this ambiguity, the population of bone-fide free-floating planetary-mass objects (PMOs) and the set of wide-orbit planets must be subsets of the $\mu$FFP population. These features of $\mu$FFPs provide limits on the low-mass end of the stellar population and the high-mass end of the planet population.

There is considerable confidence in assigning most objects to either the stellar or planetary population based on their masses. However, there is much more ambiguity for objects with masses $\sim$ 3--15 $M_{\rm Jup}$, where $M_{\rm Jup}$ is the mass of Jupiter \citep[e.g.,][and references therein]{Burrows1997,Schlaufman2018,Kirkpatrick2024}. Improving our understanding of the frequency of objects in this mass range may yield better constraints on the lowest (highest) masses generated by star (planet) formation processes. 

For stars, the microlensing stellar MF can be compared with the mass function of the lowest mass brown dwarfs derived from surveys of the nearest stellar objects \citep[e.g.,][]{Kirkpatrick2024}. For planets, the $\mu$FFP population can be compared to the wide-orbit planet populations derived from radial velocity and direct imaging observations. $\mu$FFPs also test constraints on the population of Jupiter-mass planets inferred from studies of debris disks and protoplanetary disks.

\citet{ClantonGaudi17} also tried to reconcile wide-orbit planets with the $\mu$FFP population. \citet{Sumi11} had inferred a large population of free-floating Jupiters based on an excess of short-timescale ($t_{\rm E} < 1~\mathrm{day}$) microlensing events. \citet{ClantonGaudi17} tested whether the observations could be explained by the bound planet population derived in \citet{ClantonGaudi16}  from microlensing \citep{Gould10, Sumi10}, radial velocity \citep{Montet14}, and direct imaging \citep{Lafreniere07, Bowler15} measurements of the planet frequency. The constraints available at that time from  radial velocity and direct imaging were relatively weak and limited to planets somewhat more massive than Jupiter.

Since then, the measurements of the $\mu$FFP populations have been updated several times. \citet{Mroz17FFPs} found no evidence for a large population of free-floating Jupiters in their timescale distribution, although they did find tentative evidence of a population of free-floating super-Earths. \citet{Gould22_FFPs} made an independent measurement based on the Einstein radius distribution and constrained both the frequency and power-law index of the $\mu$FFP population. \citet{Sumi23_FFPs} combined timescale and Einstein radius distributions to measure those properties for a larger sample of events. In addition, several techniques (radial velocities, microlensing, direct imaging, and coherent structures in circumstellar disks) have provided new insights into the frequency and mass distribution of planetary mass objects at 10--100~au distances from a host star \citep[e.g.,][]{Poleski21,Pearce22,Bae23_PPVII,Currie23_PPVII,Lagrange2023}.

We compare the MFs inferred from $\mu$FFPs \citep{Gould22_FFPs, Sumi23_FFPs} to the wide-orbit planet populations inferred from other techniques. We begin with a description of our methodology and then discuss the constraints on $\mu$FFPs and brown dwarfs from the microlensing perspective. After comparing these constraints to constraints from direct imaging and radial velocity surveys for massive planets, we discuss indirect constraints from structures observed in debris and protoplanetary disks. We conclude with a brief summary.

\newpage
\section{Methodology}

\subsection{Stars vs. Planets}
\begin{deluxetable}{|cccc|cc|}
\tablecaption{Mass/Spectral-Type Ranges Based on \citet{PecautMamajek13} \label{tab:st_mass}}
\tablehead{
\multicolumn{4}{|c|}{\citet{PecautMamajek13}} & \multicolumn{2}{c|}{\citet{Vigan21}}\\
\multicolumn{1}{|c}{} & \colhead{} & \multicolumn{2}{c|}{Pop. Frac} & \colhead{} &\multicolumn{1}{c|}{Pop. Frac}\\
\multicolumn{1}{|c}{Sp Type} & \colhead{Mass Range ($M_\odot$)} & \colhead{\citet{Chabrier05}} & \multicolumn{1}{c|}{\citet{Sumi23_FFPs}} & \colhead{Mass Range ($M_\odot$)} & \multicolumn{1}{c|}{\citet{Chabrier05}}}
\startdata
B & [2.43,    8.00] & 0.025 & 0.034& [1.68,   3.00] &  0.032\\
A & [1.68,    2.43] & 0.022 & 0.027&                &       \\
\hline
F & [1.07,    1.68] & 0.056 & 0.058&                &       \\
G & [0.89,    1.07] & 0.035 & 0.035& [0.58,   1.68] &  0.198\\
K & [0.58,    0.89] & 0.107 & 0.099&                &       \\
\hline
M & [0.08,    0.58] & 0.743 & 0.537& [0.30,   0.58] &  0.229\\
\enddata
\tablecomments{Population fractions normalized to the number $(0.08, 8)~M_\odot$ stars in the \citet{Chabrier05} MF.
B stars are limited to $M_* < 8~M_{\odot}$.
\citet{Vigan21} divided their stars into BA, FGK, and M groups as indicated by the horizontal lines. The values provided are the combined values for each group limited to $0.3~M_\odot < M_* <3~M_\odot$ stars.}
\end{deluxetable}

Because the $\mu$FFP population is fit as an excess to the stellar population, we can separate the microlensing MFs into stellar and planetary components. We refer to brown dwarfs as the low-mass tail of the stellar formation process and planets as any objects from the planetary component of the MF, regardless of mass.

For our quantitative comparisons, we quote the planet frequencies as number of planets per 100 stars \citep[following the conventions of ][]{Fulton21}. We define stars to be objects in the range $0.08\, M_\odot < M_* < 8\, M_\odot$. That is, we assume that neither brown dwarfs ($M < 0.08\, M_\odot$) nor high-mass stars ($M_* > 8\, M_\odot$) form a significant number of planets. We use the \citet{Chabrier05} MF as our reference stellar MF. We restrict our comparisons to objects with semi-major axis $a > 10~\mathrm{au}$  and focus on objects with masses $m_{\rm p}$ (or $m_{\rm p} \sin i$) $< 13\, M_{\rm Jup}$. We use $M_*$ for the masses of host stars, $m_{\rm p}$ for masses of the companions, and $M$ for masses of objects in a mass function.

For studies that quote planet frequencies for a particular spectral type or types rather than a host mass range, we use \citet{PecautMamajek13} to derive assumed mass ranges for a given spectral type. Table \ref{tab:st_mass}, lists the fraction of stars in each spectral type bin from the \citet{Chabrier05} MF. For studies that cover host stars from only part of the range,  we consider either the scenario that those hosts are representative of all stars or re-weight the frequencies by the fraction of $0.08~M_\odot < M_* < 8~ M_\odot$ stars represented by those hosts.

\subsection{Radial Velocity \& Direct Imaging}

For radial velocity and direct imaging, the distinction between planets and brown dwarfs is unclear, especially for objects with masses $m_{\rm p} \sim 3-15~M_{\rm Jup}$. For radial velocity studies, it may be possible to separate planets and brown dwarf companions based on mass ratio. Because direct imaging studies are still limited to detecting the most massive companions, the origin of many companions is ambiguous unless the companion lies within a disk \citep{Bowler25, Close25, Zhou25} or with additional coplanar objects \citep[e.g., the HR 8799 system][]{Marois08, Marois10} or both \citep[e.g., $\beta$ Pic b and c][]{Lagrange09, Lagrange19}.

We will compare the population of objects with masses $<13~M_{\rm Jup}$  detected by radial velocity and direct imaging to the $\mu$FFP population. If there are more radial velocity and direct imaging companions than $\mu$FFPs, we can infer the fraction  of such companions that should be classified as planetary vs. as brown dwarfs.  Having fewer radial velocity and direct imaging companions than $\mu$FFPs places a constraint on the fraction of $\mu$FFPs  that are bound vs. free-floating.

\subsection{Circumstellar Disks}

To generate the structures observed in protoplanetary and debris disks, embedded or nearby planets are one option among many \citep{Vorobyov20, Friebe22, Smallwood23, Stuber23}. Massive planets outside a disk can create spiral density waves \citep{Dong11,Cimerman21}. Inside the disk, planets can clear out material along their orbits and also generate spiral waves throughout the disk. The contrast of these features relative to the rest of the disk constrains the mass of the planet and its distance from the disk and the host star. 

In most cases, the presumed planets are at large distances from the host star; their frequency can then be directly compared with the frequency of microlensing FFPs. Most studies assume that every circumstellar disk structure (e.g., each gap in an ALMA disk) corresponds to a single planet. If there are more inferred companions than allowed by the $\mu$FFP population, this assumption is invalidated, which constrains the physics used to infer the planet properties. If there are fewer inferred companions, a significant fraction of the $\mu$FFP population can then be unbound rather than in wide-orbits.

{\section{Microlensing Constraints on the Free-Floating Planet Population}
\label{sec:ulens}}

\subsection{Overview of Microlensing FFP Detection \& Characterization}

\begin{figure}
\begin{centering}
    \includegraphics[width=0.7\textwidth]{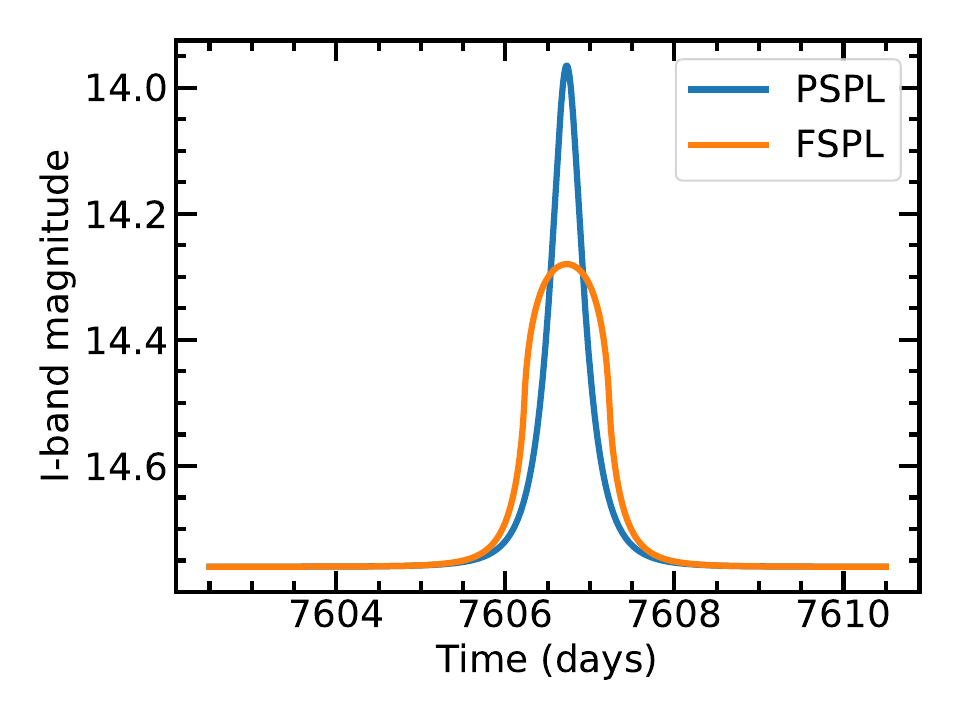}
    \caption{An example of the microlensing light curve of a $\mu$FFP based on the properties of OGLE-2016-BLG-1540 \citep{Mroz18}. The Einstein timescale of the event is $t_{\rm E} = 0.320$ day. The blue curve shows the light curve under a point-source approximation (PSPL). The blue curve includes the effect of the physical extent of the source star (FSPL), which enables a measurement of the Einstein radius.
    \label{fig:ob1540}}
\end{centering}
\end{figure}

A microlensing event occurs when a foreground object, the lens, passes in front of and magnifies the light from a (usually unrelated) background object, the source. The morphology of the resulting light curve depends on the Einstein radius, $\theta_{\rm E}$,  and Einstein timescale, $t_{\rm E}$, which are proportional to the square root of the lens mass, $M$ \citep{Paczynski86b}:
\begin{eqnarray}
    \theta_{\rm E} & = & \sqrt{\kappa\,M \pi_{\rm rel}} 
        = 2.0\, \mu\mathrm{as} \left(\frac{M}{10\, M_\oplus}\right)^{1/2} \left(\frac{\pi_{\rm rel}}{0.016\, \mathrm{mas}}\right)^{1/2} \label{eqn:theta_E}\\
    t_{\rm E} & = & \frac{\theta_{\rm E}}{\mu_{\rm rel}}
        = 0.12\, \mathrm{d} \left(\frac{M}{10\, M_\oplus}\right)^{1/2}
        \left(\frac{\mu_{\rm rel}}{6\, \mathrm{mas / yr}}\right)^{-1}
        \label{eqn:t_E}
\end{eqnarray}
where $\kappa = 8.14\, M_\odot^{-1}\, \mathrm{mas}$, $\pi_{\rm rel} \equiv \mathrm{au} (D_{\rm L}^{-1} - D_{\rm S}^{-1})$, $D_{\rm L}$ is the distance to the lens, $D_{\rm S}$ is the distance to the source, and $\mu_{\rm rel}$ is the lens-source relative proper motion. Microlensing experiments monitor the brightness of millions of stars to look for this particular light curve morphology \citep{Gaudi12}.

 Equations \ref{eqn:theta_E} and \ref{eqn:t_E} indicate that a planetary mass object will produce a microlensing event with a short timescale and will have a small $\theta_{\rm E}$ (if it is measurable). Figure \ref{fig:ob1540} shows example light curves based on the $\mu$FFP event OGLE-2016-BLG-1540 \citep{Mroz18}. Because the event was caused by a single body, it is a point-lens (PL) event, which had $t_{\rm E} = 0.32~\mathrm{day}$, $\theta_{\rm E} = 9.2~\mu\mathrm{as}$, and a Neptune-mass lens. The blue curve (PSPL)  shows the light curve assuming a point source (PS). The orange curve (FSPL) includes the finite source (FS)  effect (i.e., accounting for the physical extent of the source star), which is one way to measure $\theta_{\rm E}$ \citep{Yoo04_CMDMethod}.

A microlensing event that appears as a single lens event due to a planetary-mass object may lack a microlensing signature due to the host star. Given a detected planet, the probability of detecting a host star decreases rapidly with the orbital separation of the planet.  For a planet that appeared to be separated from its host star by $6.6~\theta_{\rm E}$ (analogous to a Neptune-like orbit),  there is only a $\sim 20\%$ probability of detecting the microlensing signature of the host star \citep{Gould16, ClantonGaudi17}. For individual $\mu$FFP candidates, the microlensing light curve is usually only sufficient to rule out host stars within $\sim 10~\mathrm{au}$ \citep{Mroz18, Mroz20FFP, Ryu21_kb2820},  making it ambiguous whether any given $\mu$FFP candidate is truly free-floating or merely on a wide orbit. 

High-resolution follow-up observations can be used to search for potential host stars, but such searches are still in the early stages. For example, \citet{Mroz24} searched for hosts for five $\mu$FFP candidates, but were only able to rule out the most massive $\sim$10--35\% of possible host stars.

Even without knowing whether or not there are host stars for individual objects, the $\mu$FFP population measures the combined sum of the wide-orbit planet population and the population of free-floating PMOs. For example, if all $\mu$FFPs are wide orbit planets, there cannot be any true FFPs 
and vice versa. Based on published limits on host stars for individual events, we set $10~\mathrm{au}$ as the lower bound of the semi-major axis range probed by $\mu$FFPs.

 Equations \ref{eqn:theta_E} and \ref{eqn:t_E} also show that  while short $t_{\rm E}$  or small $\theta_{\rm E}$ events are strong candidates for having planetary mass lenses, the mass of any given object is ambiguous.  Hence, the properties of the $\mu$FFP  population are inferred by measuring statistically significant excesses to the $t_{\rm E}$  or  $\theta_{\rm E}$  distributions after accounting for the stellar population. The first $\mu$FFP population studies measured or constrained excesses to the $t_{\rm E}$  distribution \citep{Sumi11, Mroz17FFPs}. The studies discussed in this work have measured the $\theta_{\rm E}$ distribution, which has fewer unknowns but is restricted to a smaller sample of events. Relatively few objects with small $\theta_{\rm E}$ have been detected \citep{Mroz18,Mroz19,Mroz20_FFP_0551,Mroz20FFP,Kim21_kb2073,Ryu21_kb2820,Koshimoto23_FFPs,Jung24}. 
 
\subsection{Measurements of the Microlensing FFP Population}

Recently, both the Korea Microlensing Telescope Network \citep[KMTNet;][]{Kim16_KMTNet} and Microlensing Observations in Astrophysics \citep[MOA; ][]{Bond04} surveys have made measurements of the $\mu$FFP MF using samples that include events with measured $\theta_{\rm E}$.

\citet{Gould22_FFPs} measured a power-law distribution for $\mu$FFPs from a statistical sample of 30 events with measured $\theta_{\rm E}$, including 4 detections in the ``planetary" regime ($\lesssim10\, \mu$as in their case). In particular, they observed an absence of objects with $9\, \mu$as $< \theta_{\rm E} < 26\, \mu$as, which they dubbed the ``Einstein Gap." Because of this gap, they suggested that the MF could be comprised of two separate MF, one for ``planets" and one for ``stars".  For the planetary component, they derive
\begin{equation}
\frac{dN}{ d \log M} = \frac{0.28\pm0.09}{\mathrm{dex}\times {\rm stars}} \times \left(\frac{M}{38\,M_{\oplus}}\right)^{-p}
\label{eqn:gould}
\end{equation}
with $p$ likely to be in the range $0.9 < p < 1.2$; $p = 0.6$ is ruled out.

\citet{Gould22_FFPs} showed that the $\mu$FFP population is consistent with  the short timescale events from \citet{Mroz17FFPs} and with the bound microlensing planet population from \citet{Poleski21}, who measured the occurrence rate of planets from $~5-15~\mathrm{au}$  to be $1.4^{+0.9}_{-0.6}$  per star with a power law index of $p\sim 1.0$. \citet{Gould22_FFPs} also found consistency between their population and limits on the population of `Oumuamua-like  objects and the fraction of Uranus and Neptune-like objects that would be detected as free-floating vs. bound planets.

\citet{Sumi23_FFPs} conducted a joint  statistical analysis of events from the MOA survey \citep{Koshimoto23_FFPs}, some with measured $\theta_{\rm E}$ and others with  only $t_{\rm E}$. They derived a similar power-law mass function to \citet{Gould22_FFPs}:
\begin{equation}
\frac{dN}{d \log M} = \frac{2.18^{+0.52}_{-1.40}}{\mathrm{dex}\times {\rm stars}}\times \left(\frac{M}{8\, M_\oplus}\right)^{-p} \quad ;\quad p = 0.96^{+0.47}_{-0.27} .
\end{equation}

\begin{figure}
    \includegraphics[width=\textwidth]{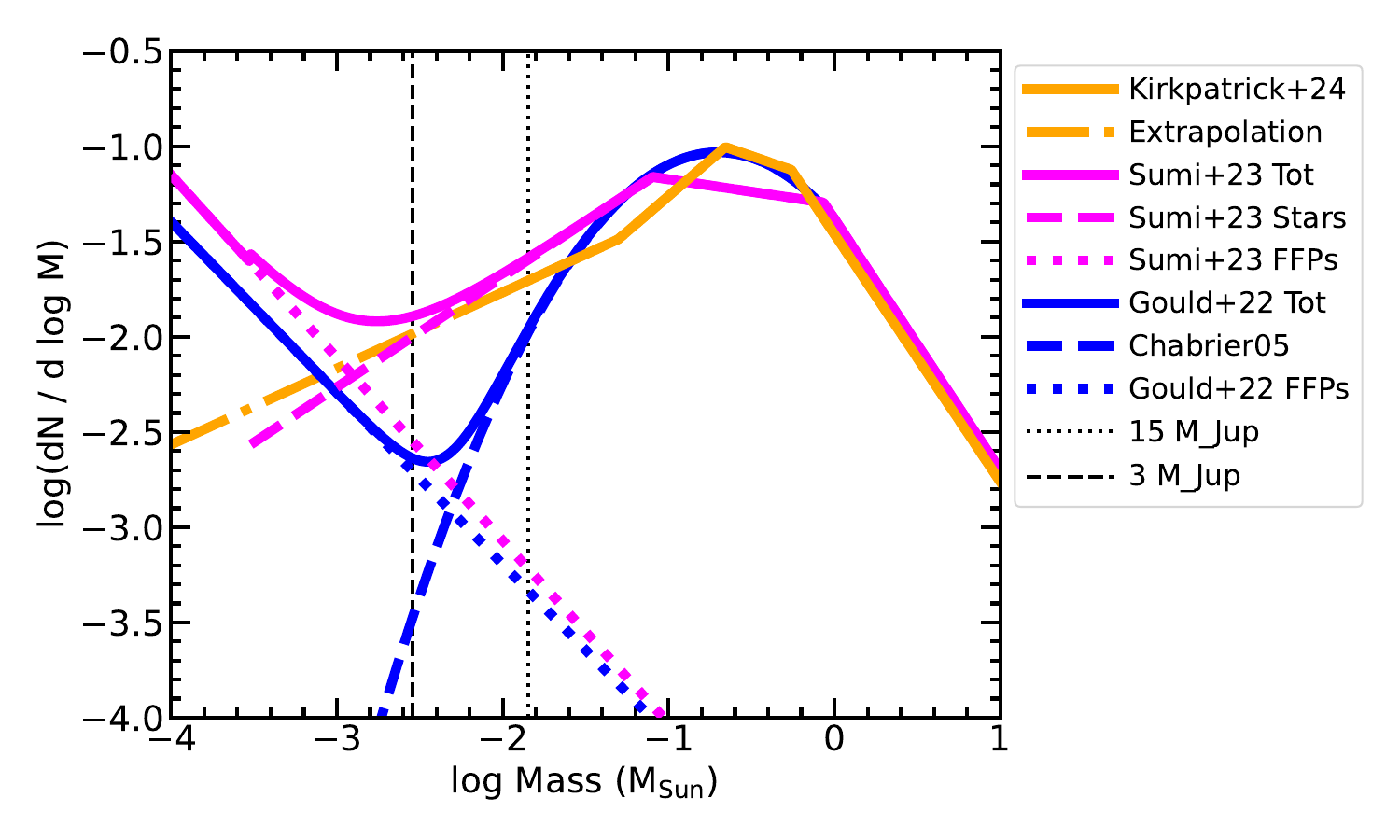}
    \caption{Mass functions from \citet{Gould22_FFPs} and \citet{Chabrier05}, \citet{Sumi23_FFPs} (magenta), and \citet{Kirkpatrick2024} (orange) The dashed lines show the FFP component while the dotted lines show the stellar component. The solid line shows their sum. The orange dash-dot line indicates the \citet{Kirkpatrick2024} MF is extrapolated beyond the available data below $10^{-2}~M_\odot$. The black, vertical, dashed and dotted lines are drawn at $3\,M_{\rm Jup}$ and $15\,M_{\rm Jup}$, respectively. Although this regime is often considered to be ``planetary" (as defined by mass), in these mass functions, the ``stellar" component is clearly dominant. \label{fig:mfs}}
\end{figure}

Figure \ref{fig:mfs} compares the MF from \citet{Gould22_FFPs} and \citet{Sumi23_FFPs}. Both studies quote their MFs in planets per star. \citet{Sumi23_FFPs} defines stars as objects from their stellar MF with $3\times10^{-4}~M_\odot < M < 8~M_\odot$. We renormalize the \citet{Sumi23_FFPs} MF to have equal numbers of stars in this range as in the \citet{Chabrier05} MF. \citet{Gould22_FFPs}  does not specify a particular mass range as corresponding to stars, so we use the same range as for \citet{Sumi23_FFPs} to renormalize the MF. Because stellar MFs are steep at both low and high masses, the normalization is relatively insensitive to the choices of the upper and lower limits. Using $0.013 ~M_\odot$ as the lower limit or $100~M_\odot$ as the upper limit changes the normalization by $<1\%$.

\subsection{Comparison of Sumi et al (2023) Stellar MF With Other MFs}

The stellar MF of \citet{Sumi23_FFPs} consists of a 3-part power law and
differs from \citet{Chabrier05} below $\sim 1~M_{\odot}$. Figure \ref{fig:mfs} shows that \citet{Sumi23_FFPs} predicts fewer M dwarfs and significantly more brown dwarfs than \citet{Chabrier05}, but the MFs agree well for higher-mass stars. These features are quantified in spectral-type bins in Table \ref{tab:st_mass}. For comparison, we also plot the IMF from \citet{Kirkpatrick2024} in Figure \ref{fig:mfs}. This comparison shows that the abundance of brown dwarfs identified in the \citet{Sumi23_FFPs} MF is consistent with  observations of the MF in the Solar neighborhood. All three stellar MFs are reasonably consistent with each other except in the brown dwarf regime. 

We choose to quote all of our planet frequencies, including those for the \citet{Sumi23_FFPs} $\mu$FFP MF, relative to the number of $0.08~M_\odot < M < 8~M_\odot$ stars from the \citet{Chabrier05} MF. If we were to use the \citet{Sumi23_FFPs} stellar MF instead, there would be $\sim 20\%$ fewer $0.08~M_\odot < M < 8~M_\odot$ stars or $\sim25\%$ more planets per star.

\newpage
\subsection{Comparison of Gould et al (2022) \& Sumi et al (2023) Microlensing FFP MFs}

\begin{deluxetable}{l|rrr|rrr}
\tablecaption{Number of Objects per 100 Stars \label{tab:ffp_freq}}
\tablehead{
\multicolumn{1}{c|}{} & \multicolumn{3}{c|}{\citet{Gould22_FFPs}} & \multicolumn{3}{c}{\citet{Sumi23_FFPs}}\\
\multicolumn{1}{c|}{Mass Range} &
\multicolumn{1}{c}{Total} & \colhead{FFPs} & \multicolumn{1}{c|}{Stars} &\multicolumn{1}{c}{Total} & \colhead{FFPs} & \multicolumn{1}{c}{Stars}}
\startdata
Convenient Divisions: & \multicolumn{3}{c|}{} & \multicolumn{3}{c}{}\\
$[1\,M_{\rm Jup}, 13\,M_{\rm Jup}]$     &    4.2 &    2.4 &    1.9&   17.4 &    3.4 &   13.9\\
$[1\,M_{\rm Sat},  1\,M_{\rm Jup}]$     &    5.1 &    5.1 &    0.0&   10.2 &    8.2 &    2.0\\
$[1\,M_{\rm Nep},  1\,M_{\rm Sat}]$     &   28.4 &   28.4 &    0.0&   50.1 &   50.1 &    0.0\\
$[1\,M_{\rm Earth}, 1\,M_{\rm Nep}]$    &  430.9 &  430.9 &    0.0&  888.1 &  888.1 &    0.0\\
\hline
``Giant" Planets: & \multicolumn{3}{c|}{} & \multicolumn{3}{c}{}\\
$[1\,M_{\rm Sat}, 13\,M_{\rm Jup}]$     &    9.3 &    7.5 &    1.9&   27.6 &   11.7 &   15.9\\
$[1\,M_{\rm Nep}, 13\,M_{\rm Jup}]$     &   37.8 &   35.9 &    1.9&   77.7 &   61.8 &   15.9\\
$[1\,M_{\rm Nep},  1\,M_{\rm Jup}]$     &   33.6 &   33.6 &    0.0&   60.3 &   58.3 &    2.0\\
\hline
Comparisons to Other Work: & \multicolumn{3}{c|}{} & \multicolumn{3}{c}{}\\
$[13\,M_{\rm Jup}, 80\,M_{\rm Jup}]$    &   27.8 &    0.2 &   27.6&   33.7 &    0.1 &   33.6\\
$[5\,M_{\rm Jup}, 13\,M_{\rm Jup}]$     &    2.0 &    0.4 &    1.6&    8.1 &    0.5 &    7.7\\
$[1\,M_{\rm Jup}, 75\,M_{\rm Jup}]$     &   30.0 &    2.6 &   27.4&   49.2 &    3.6 &   45.6\\
$[1\,M_{\rm Jup}, 20\,M_{\rm Jup}]$     &    6.6 &    2.4 &    4.2&   22.6 &    3.6 &   19.0\\
\enddata
\tablecomments{Normalized to the number $(0.08, 8)\, M_\odot$ stars in the \citet{Chabrier05} MF.}
\end{deluxetable}

Figure \ref{fig:mfs} shows the \citet{Sumi23_FFPs} $\mu$FFP MFs, and Table \ref{tab:ffp_freq} presents integrated frequencies over various mass ranges. We list the total number of objects in each mass range, as well as separating them into the planetary and stellar components. For simplicity, we do not include uncertainties in the MFs in our calculations: for \citet{Gould22_FFPs}, we assume $p=0.9$ and for \citet{Sumi23_FFPs}, we calculate frequencies for $p=0.96$. 

While the two MFs have a similar slope, the estimates of the $\mu$FFP frequencies from \citet{Gould22_FFPs} are \num{a factor of $\sim1.5$ to 2} smaller than the values derived from \citet{Sumi23_FFPs}. However, given the Poisson uncertainties in the \citet{Gould22_FFPs} measurement (4 observed FFPs), these values are roughly consistent. Even with these uncertainties, Table \ref{tab:ffp_freq} indicates there are multiple small planets ($<1\,M_{\rm Nep}$) per star in wide orbits or ejected from their parent systems, whereas there is \num{ $< 1$} wide-orbit planet or free-floating PMO with a mass in the range $(1\,M_{\rm Nep}, 13\,M_{\rm Jup})$ per star. 

{\section{Comparison to Other Measurements of the Wide-Orbit Planet Population}
\label{sec:comparisons_wop}}

{\subsection{Direct Imaging} \label{sec:di}}

Attempts to detect exoplanets with direct imaging benefit from increasingly sophisticated and sensitive AO cameras and high quality software \citep[cf.][]{Currie23_PPVII}. However, direct imaging has only been sensitive to the very highest mass end of the ``planetary" mass  regime: $m_{\rm p} > 1$ to a few $~M_{\rm Jup}$. In microlensing populations, the planet and brown dwarf populations overlap at these masses (Fig. \ref{fig:mfs}). We can test whether the direct imaging planet population is consistent with being planetary by seeing whether it can be explained by the $\mu$FFP population or whether it also requires contributions of objects from the stellar population.

\vspace{12pt}
{\subsubsection{Nielsen et al (2019)}
\label{sec:nielsen}}

\citet{Nielsen19} analyze GPI data for 300 stars, which yielded 6 planets and 3 brown dwarfs. The host stars range in mass from $0.2\, M_\odot$ to $5\, M_{\odot}$.
\citet{Nielsen19} fit  their population with  a power law: 
\begin{equation}
d^2N/(dm_{\rm p}~da) \propto m_{\rm p}^{\alpha} ~ a^{\beta}.
\end{equation} 
For the full host sample, they derive $\alpha = -2.3^{+0.8}_{-0.7}$. This power-law index ($\alpha \equiv -(p + 1)$) is consistent with the values for the microlensing FFP population.

\citet{Nielsen19} infer raw frequencies of \num{2.1--5.4} (68\% confidence interval) 5--13 $M_{\rm Jup}$ companions per 100  stars with $0.2\, M_\odot < M_* < 5\, M_{\odot}$. For comparison, \citet{Gould22_FFPs} calculate the total frequency of 5--13 $M_{\rm Jup}$ objects as \num{2.0 per 100 stars} from a combination of FFPs and brown dwarfs. From \citet{Sumi23_FFPs}, we infer a total frequency of \num{8.1 objects per 100 stars} with the majority coming from the tail of the stellar population (see Table \ref{tab:ffp_freq}). If we assume the results for $0.2\, M_\odot$ to $5\, M_{\odot}$ stars are representative of all stars, the \citet{Nielsen19} frequencies are only consistent with the microlensing population if most of the companions are drawn from the brown dwarf population.

However, stars from $0.2\, M_\odot$ to $5\, M_{\odot}$ account for only \num{$\sim 65$\%} of the stellar population. If we assume that no stars outside of this mass range have 5--13 $M_{\rm Jup}$ companions, the frequency is reduced to \num{$2.3^{+ 1.2}_{-0.9}$ planets per 100 stars}. This result is still in tension with the hypothesis that all of these companions are planets. 

One way to reconcile these frequencies would be to use the lower limit from \citet{Nielsen19} from the 95\% confidence interval: \num{1.1 planets per 100 stars}. Combined with the correction factor for the fraction of hosts and the uncertainties in the $\mu$FFP population, this may be enough to bring the measurements into agreement. But, it also requires that {\it all} of the $\mu$FFPs in this mass range are bound, wide-orbit planets.

All of the planets in \citet{Nielsen19} orbit stars with $M_* > 1.5 M_{\odot}$ ($\sim 40\%$ of their sample), so they also analyze this subsample of stars. They find a similar power-law index to the full sample and a frequency of 5--13 $M_J$ companions at 10--100~au to be \num{$9^{+5}_{-4}$ planets} per 100 stars with $1.5~M_\odot < M_* < 5~M_\odot$. Such a large frequency of planets, if it extended to host stars of all masses, is inconsistent with the $\mu$FFP population. However, 1.5--5 $M_\odot$ stars account for only about \num{$\sim 6\%$}  of the stellar population. Assuming no other stars have companions of this mass, which is supported by the lack of detections around lower-mass stars, the planet frequency from \citet{Nielsen19} becomes \num{$0.5^{+ 0.3}_{- 0.2}$ planets per 100 stars}. This frequency is \num{in good agreement} with the $\mu$FFP population but still requires all $\mu$FFPs to be bound planets.

\citet{Nielsen19} also measure the frequency of companions in the 13--80 $M_{\rm Jup}$ range (the commonly used mass range for brown dwarfs). After scaling according to the host-star population (0.2--5.0 $M_\odot$),  the frequency of such objects is \num{$0.5^{+ 0.5}_{- 0.3}$ objects per 100 stars}. These values are roughly consistent with the \citet{Gould22_FFPs} and \citet{Sumi23_FFPs} $\mu$FFP populations (\num{0.2 and 0.1 planets per 100 stars},  respectively). It also suggests that the vast majority of microlensing objects in this mass range cannot be bound to a star, which is consistent with the hypothesis that they are part of the tail of the stellar MF. 

\vspace{12pt}
{\subsubsection{Vigan et al (2021)}
\label{sec:vigan}}

\citet{Vigan21} analyze SPHERE results for 105~stars and are sensitive
to 1--$75\, M_{\rm Jup}$ companions over the semi-major axis range $a$ = 5--300~au. They conclude that the predicted fraction of the population due to planets vs. brown dwarfs is a function of stellar mass: a large fraction of substellar companions are drawn from the planetary MF for BA stars, while for M stars, most companions are drawn from the stellar MF. \citet{Vigan21} report their results as the frequency of systems, which represents a lower limit on the frequency of planets, if stars can host multiple planets (as in the case of HR8799). 

\citet{Vigan21} derive the frequencies of at least one
substellar ($1~M_{\rm Jup} < m_{\rm p} < 75~M_{\rm Jup}$) companion as \num{23.0$^{+13.5}_{-9.7}$} (BA stars), 
\num{5.8$^{+4.7}_{-2.8}$} (FGK stars), 
and \num{12.6$^{+12.9}_{-7.1}$} (M stars) per 100 stars. 

To obtain the total companion frequency, we re-weight each spectral-type bin based on the fraction of the stellar population it corresponds to (Table \ref{tab:st_mass}). The mass ranges for the M and BA bins are truncated because \citet{Vigan21} state their sample is limited to host stars with $0.3~M_\odot < M_* < 3~M_{\odot}$. 

After correcting for these population fractions and assuming no stars outside the range $0.3~M_\odot < M_* < 3~M_{\odot}$ have companions, the \citet{Vigan21} results imply a total frequency of \num{$2.0^{+2.7}_{-1.4}$ substellar companions per 100 stars}. These values are comparable to the expected FFP population from microlensing: \num{2.6 planets per 100 stars} \citep{Gould22_FFPs} and \num{3.6 planets per 100 stars} \citep{Sumi23_FFPs}. See Table \ref{tab:ffp_freq}. 
However, as with \citet{Nielsen19}, if we assume that all of these companions correspond to the $\mu$FFPs, then all of the $\mu$FFPs must be bound. If we consider the \citet{Vigan21} sample to be representative of all stars (instead of excluding planets around stars with $M_* < 0.3~M_\odot$ or $M_* > 3~M_{\odot}$), the frequencies increase by a factor of 2.

Similar to microlensing studies, \citet{Vigan21} consider their direct imaging detections as a mix of planets and brown dwarfs.
We scale the parameterized population models from \citet{Vigan21}
by the host star population fraction and assume that no stars with $M_* < 0.3~M_\odot$ or $M_* > 3~M_{\odot}$ have companions. The total frequencies are \num{$\ge 0.5^{+ 3.4}_{- 0.3}$ planets per 100 stars} and \num{$\ge2.0^{+ 2.7}_{- 1.4}$ brown dwarfs per 100 stars}. Even with the factor of 2 correction, these frequencies are below the limits inferred from $\mu$FFPs and the microlensing stellar MF, respectively. Because the microlensing stellar population is much larger than the \citet{Vigan21} brown dwarf sample, this result implies that a significant fraction of the microlensing objects either do not have host stars or orbit stars with masses outside the range $0.3~M_{\odot} < M_* < 3~M_{\odot}$ (see Table \ref{tab:ffp_freq}). 

The one caveat is that \citet{Vigan21} assumed fixed power-law MFs with $\beta = -1.31$ (where $\beta \equiv -(p+1)$) for the planetary MF, which is shallower than allowed for the microlensing FFP population \citep{Gould22_FFPs}. A steeper power-law would decrease the number of larger mass objects in the planetary component of the \citet{Vigan21} population, but the number of observed objects and total frequencies remain fixed. This alternative implies a larger frequency of brown dwarfs and smaller frequency of planets than considered above; however, the results remain consistent with the microlensing populations.

\newpage
{\subsection{Radial Velocity: Fulton et al (2021)} \label{sec:rv}}

\begin{figure}
    \centering
    \includegraphics[height=0.4\textheight]{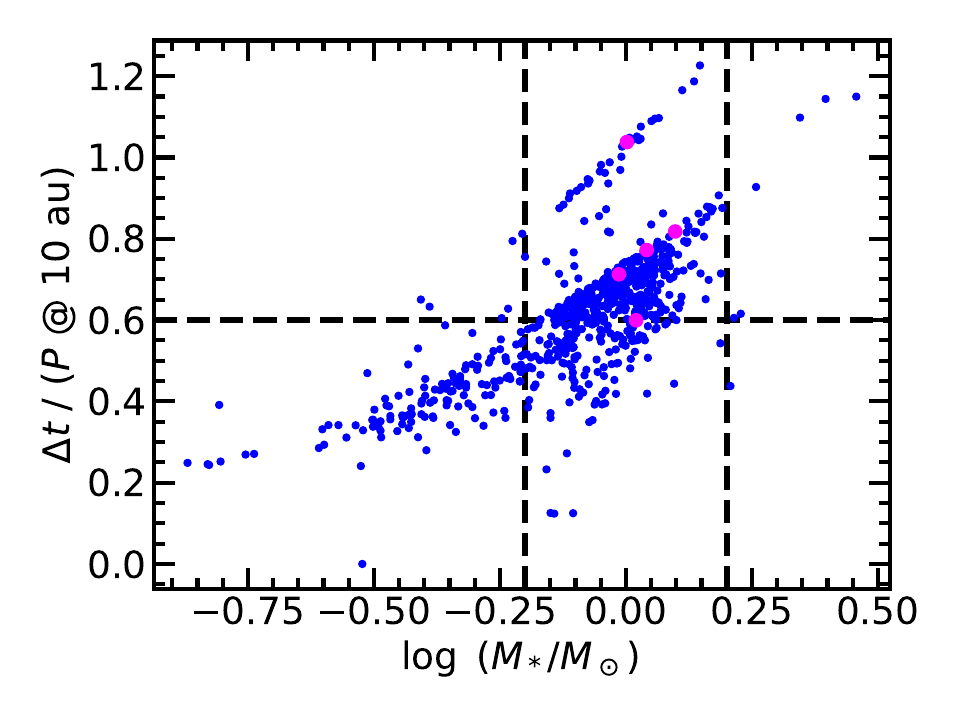}
    \caption{Comparison of host mass ($\log M$) and the duration of observations $\Delta t$ as a fraction of the period $P$ at $a = 10~\mathrm{au}$ for the radial velocity sample from \citet{Rosenthal21} and \citet{Fulton21}. Blue points are for all stars. Magenta points indicate stars with planets with $a > 10~\mathrm{au}$. The horizontal dashed line at $y=0.6$ is roughly the minimum for detected $a > 10~\mathrm{au}$ planets. The vertical dashed lines are drawn at $\log M = -0.2, 0.2$. 
    \label{fig:rosenthal}}
\end{figure}

Radial velocity planet searches offer another technique to probe the wide-orbit planet population via direct detections. Several recent radial velocity measurements of the giant planet population ``beyond the ice line" \citep{Fernandes19, Wittenmyer20, Fulton21} indicate that the frequency of gas giants appears to increase with $a$ out to 3--10~au and then drop. Of these, only \citet{Fulton21} claims sensitivity to planets beyond 10~au, so we focus our comparison on those results.

\citet{Fulton21} analyze a sample of 719 stars containing 178 planets from the California Legacy Survey \citep{Rosenthal21}. Based on their figure 9, we estimate the frequency of $1$ -- $20\, M_{\rm Jup}$ objects in orbits from 10--50~au as \num{$6.3_{-1.9}^{+3.0}$ planets per 100 stars} (median and 68\% confidence interval), and the frequency of $5\, M_{\rm Jup}$ -- $13\, M_{\rm Jup}$ objects in orbits from 10--100~au is \num{$3.5_{-1.6}^{+2.3}$ planets per 100 stars}. The values for the $\mu$FFP population from Table \ref{tab:ffp_freq} for these mass ranges lie at the extreme edge of the probability distributions in figure 9 from \citet{Fulton21}, indicating the measurements are inconsistent.

\citet{Fulton21} describe the sample as consisting of FGKM stars, corresponding to most stars; renormalizing the values to account for stars with $M_* > 1.68~M_\odot$ has minimal effect on the totals. However, the sample has low completeness beyond $10~\mathrm{au}$. Because of Kepler's third law, the completeness could be correlated with host mass. To test this possibility, we use the data from \citet{Rosenthal21} to calculate the total duration of observations, $\Delta t$, for each star. We divide this by the period, $P$, at $10~\mathrm{au}$ for that star to determine the fraction of the orbital period that has been observed. Figure \ref{fig:rosenthal} shows that there is indeed a strong correlation between this fraction and the host mass.

Figure \ref{fig:rosenthal} also highlights the five companions in the sample with $a > 10~\mathrm{au}$ and $m \le 13~M_{\rm Jup}$. The minimum value of $\Delta t / (P~@~10~\mathrm{au})$ is \num{$\sim 0.6$}. If we assume the sample is only complete for stars above this value, then the range of host masses probed by \citet{Fulton21} for wide-orbit planets is approximately \num{$-0.2 < \log M_* / M_\odot < 0.2$} or \num{$0.63~M_\odot < M_\star < 1.58~M_\odot$}. Stars in this range account for only \num{17\%} of all stars.

Applying this correction factor to the \citet{Fulton21} frequencies, yields $1.1^{+ 0.5}_{- 0.3}$ objects with masses $1~M_{\rm Jup} < m < 20~M_{\rm Jup}$ in orbits from 10--50~au per 100 stars and 
$0.6^{+ 0.4}_{- 0.3}$ objects with masses $5~M_{\rm Jup} < m_{\rm p} < 13~M_{\rm Jup}$ in orbits from 10--100~au per 100 stars. These values are consistent with the $\mu$FFP population. But for the $5~M_{\rm Jup} < m_{\rm p} < 13~M_{\rm Jup}$ bin, if all such objects are planets, all $\mu$FFPs in this mass range must be bound. 
Also, no technique should detect planets of this mass around stars with masses outside the range $0.63~M_\odot < M_* < 1.58~M_\odot$ because they should not exist. If they do exist, a significant fraction of the \citet{Fulton21} objects are brown dwarfs rather than planets.

\vspace{12pt}
{\section{Comparison to Planets inferred from Disk Structures}
\label{sec:comparisons_disks}}

{\subsection{Debris Disks: Pearce et al (2022)}\label{sec:disks}}

\begin{figure}
\begin{center}
    \includegraphics[height=0.8\textheight]{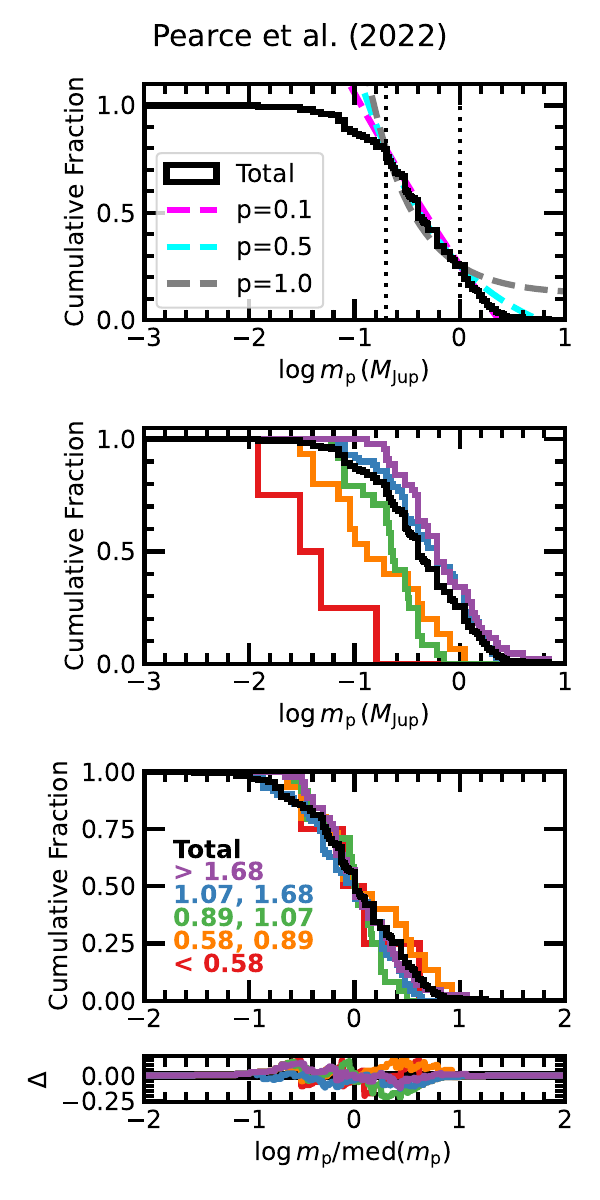}
    \caption{Cumulative distributions of minimum planet masses from \citet{Pearce22} assuming the inner disk edge is sculpted by a single planet. The distribution for the full sample is shown in black, while the subsamples for different spectral types are shown in color (red=M, blue=K, green=G, blue=F, purple=A/B). The top and middle panels show the normalized cumulative distributions. Power-laws with various indices ($dN/d\log M_p \propto M_p^{-p}$) are shown in the top panel, normalized to fit the observed distribution at the points indicated by the dotted lines. The bottom panels show the inferred debris-disk planet distributions shifted by the median mass for each group and the difference of the shifted distributions relative to the full sample. \label{fig:pearce}}
\end{center}
\end{figure}

\begin{figure}
\centering
    \includegraphics[height=0.4\textheight]{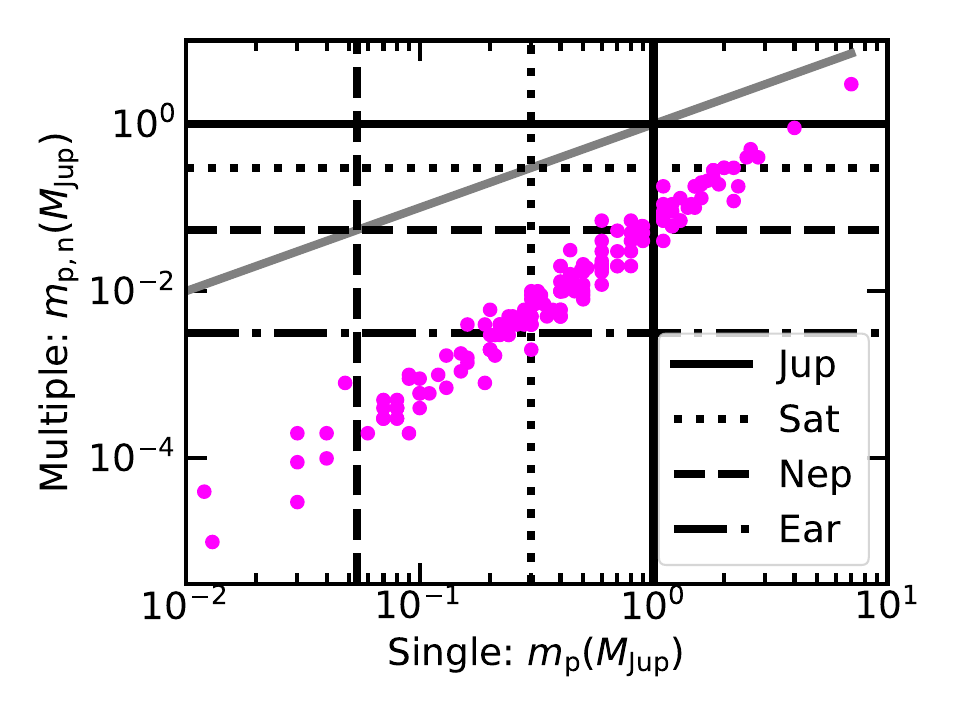}
    \caption{Magenta points compare the minimum masses calculated by \citet{Pearce22} for a single planet to shape the disk ($m_{\rm p}$, x-axis) vs. multiple, equal-mass planets ($m_{\rm p, n}$, y-axis). Gray solid line is 1-to-1. Black lines show masses of solar system planets (see legend). 
    \label{fig:pearce_Mpn}}
\end{figure}

\begin{table}
\footnotesize
\caption{Planets per 100 Stars of a given Spectral Type from \citet{Pearce22}\label{tab:pearce_spt}}
\begin{tabular}{lrrrrr}
\hline\hline
SpT      & $\min M_*$ & $\max M_*$ & $N_{\rm stars}$ & Pop Frac. & DD Frac.\\
M        &  0.22 &  0.58 &       4 &    0.355 &     0.25\\
\hline
Planet Mass Range                        &  $N_{\rm pl}$ &   frac &     All & DD-only\\
\hline
$[1\,M_{\rm Jup}, 13\,M_{\rm Jup}]$      &             0 &  0.000 &     0.0 &    0.0\\
$[1\,M_{\rm Sat},  1\,M_{\rm Jup}]$      &             0 &  0.000 &     0.0 &    0.0\\
$[1\,M_{\rm Nep},  1\,M_{\rm Sat}]$      &             1 &  0.250 &     8.9 &    2.2\\
$[1\,M_{\rm Earth}, 1\,M_{\rm Nep}]$     &             3 &  0.750 &    26.6 &    6.7\\
                                  Total: &             4 &  1.000 &    35.5 &    8.9\\
\hline
\hline
SpT      & $\min M_*$ & $\max M_*$ & $N_{\rm stars}$ & Pop Frac. & DD Frac.\\
K        &  0.58 &  0.89 &      17 &    0.107 &     0.25\\
\hline
Planet Mass Range                        &  $N_{\rm pl}$ &   frac &     All & DD-only\\
\hline
$[1\,M_{\rm Jup}, 13\,M_{\rm Jup}]$      &             1 &  0.059 &     0.6 &    0.2\\
$[1\,M_{\rm Sat},  1\,M_{\rm Jup}]$      &             5 &  0.294 &     3.1 &    0.8\\
$[1\,M_{\rm Nep},  1\,M_{\rm Sat}]$      &             6 &  0.353 &     3.8 &    0.9\\
$[1\,M_{\rm Earth}, 1\,M_{\rm Nep}]$     &             3 &  0.176 &     1.9 &    0.5\\
                                  Total: &            15 &  0.882 &     9.4 &    2.4\\
\hline
\hline
SpT      & $\min M_*$ & $\max M_*$ & $N_{\rm stars}$ & Pop Frac. & DD Frac.\\
G        &  0.89 &  1.07 &      26 &    0.035 &     0.25\\
\hline
Planet Mass Range                        &  $N_{\rm pl}$ &   frac &     All & DD-only\\
\hline
$[1\,M_{\rm Jup}, 13\,M_{\rm Jup}]$      &             0 &  0.000 &     0.0 &    0.0\\
$[1\,M_{\rm Sat},  1\,M_{\rm Jup}]$      &            10 &  0.385 &     1.3 &    0.3\\
$[1\,M_{\rm Nep},  1\,M_{\rm Sat}]$      &            14 &  0.538 &     1.9 &    0.5\\
$[1\,M_{\rm Earth}, 1\,M_{\rm Nep}]$     &             0 &  0.000 &     0.0 &    0.0\\
                                  Total: &            24 &  0.923 &     3.2 &    0.8\\
\hline
\hline
SpT      & $\min M_*$ & $\max M_*$ & $N_{\rm stars}$ & Pop Frac. & DD Frac.\\
F        &  1.07 &  1.68 &      82 &    0.056 &     0.50\\
\hline
Planet Mass Range                        &  $N_{\rm pl}$ &   frac &     All & DD-only\\
\hline
$[1\,M_{\rm Jup}, 13\,M_{\rm Jup}]$      &            24 &  0.293 &     1.6 &    0.8\\
$[1\,M_{\rm Sat},  1\,M_{\rm Jup}]$      &            28 &  0.341 &     1.9 &    1.0\\
$[1\,M_{\rm Nep},  1\,M_{\rm Sat}]$      &            18 &  0.220 &     1.2 &    0.6\\
$[1\,M_{\rm Earth}, 1\,M_{\rm Nep}]$     &             0 &  0.000 &     0.0 &    0.0\\
                                  Total: &            70 &  0.854 &     4.8 &    2.4\\
\hline
\hline
SpT      & $\min M_*$ & $\max M_*$ & $N_{\rm stars}$ & Pop Frac. & DD Frac.\\
Massive  &  1.68 &  3.70 &      49 &    0.039 &     0.50\\
\hline
Planet Mass Range                        &  $N_{\rm pl}$ &   frac &     All & DD-only\\
\hline
$[1\,M_{\rm Jup}, 13\,M_{\rm Jup}]$      &            15 &  0.306 &     1.2 &    0.6\\
$[1\,M_{\rm Sat},  1\,M_{\rm Jup}]$      &            22 &  0.449 &     1.8 &    0.9\\
$[1\,M_{\rm Nep},  1\,M_{\rm Sat}]$      &             7 &  0.143 &     0.6 &    0.3\\
$[1\,M_{\rm Earth}, 1\,M_{\rm Nep}]$     &             0 &  0.000 &     0.0 &    0.0\\
                                  Total: &            44 &  0.898 &     3.5 &    1.8\\
\hline\hline
\end{tabular}
\tablecomments{The header of each section gives the properties of the host stars. $N_{\rm pl}$ is the number of $a > 10~\mathrm{au}$ in a particular mass range and, ``frac" is $N_{\rm pl}$ divided by the total number of such planets at all semi-major axes. Because some of the inferred objects have $m_{\rm p} > 
13~M_{\rm Jup}$, the ``frac" total may be $<1$.``All" is the number of objects per 100 stars assuming the \citet{Pearce22} sample represents the planet population for all stars. ``DD-only" assumes that the sample represents only those with debris disks, i.e., other stars do not have planets.}
\end{table}

\begin{deluxetable}{lrrrrr}
\tablecaption{Total Planets per 100 Stars from \citet{Pearce22} \label{tab:pearce_all}}
\tablehead{
\colhead{Planet Mass Range} & \colhead{$N_{\rm pl}$} &\colhead{frac} & \colhead{All} & \colhead{DD-only}
}
\startdata
$[1\,M_{\rm Jup}, 13\,M_{\rm Jup}]$  &  40 &  0.225 &     3.5 &     1.6 \\
$[1\,M_{\rm Sat},  1\,M_{\rm Jup}]$  &  65 &  0.365 &     8.1 &     3.0 \\
$[1\,M_{\rm Nep},  1\,M_{\rm Sat}]$  &  46 &  0.258 &    16.3 &     4.5 \\
$[1\,M_{\rm Earth}, 1\,M_{\rm Nep}]$ &   6 &  0.034 &    28.5 &     7.1 \\
\hline
$[5\,M_{\rm Jup}, 13\,M_{\rm Jup}]$  &   1 &  0.006 &     0.1 &     0.0 \\
\enddata
\tablecomments{$N_{\rm pl}$ is the number of $a > 10~\mathrm{au}$ in a particular mass range and, `frac' is $N_{\rm pl}$ divided by the total number of such planets at all semi-major axes. `All' is the number of objects per 100 stars assuming the \citet{Pearce22} sample represents the planet population for all stars. `DD-only' assumes that the sample represents only those with debris disks, i.e., other stars do not have planets.}
\end{deluxetable}

Although many direct imaging surveys do not yield a planet, they often
detect scattered light from a debris disk \citep[e.g.,][]{Esposito20}. These detections provide one way to indirectly infer the presence of planets and their properties. \citet{Stuber23} demonstrate that a planet orbiting inside a debris disk sculpts the inner disk in a way that allows observations to infer the mass of the planet.  \citet{Friebe22} also illustrate gap structures imposed on a debris disk by migrating planets of different masses. Planets may also induce other kinds of disk structures such as warps or spiral density waves.

\citet{Pearce22} use the morphologies of debris disks around 178 host stars to constrain the masses of the planets assumed to be sculpting them. They consider the minimum planet required to stir the disk and maintain its inner edge at a fixed distance from the host star. Constraints for maintaining the inner edge of the disk are stronger; we use these estimates for our comparison of planet populations. 

Figure \ref{fig:pearce} shows the cumulative distribution of the population of planets from \citet{Pearce22} assuming the inner edge of the disk is sculpted by a single planet. We restrict the planet population to those planets with semi-major axes \num{$a > 10$ au} (\num{158} planets with a maximum semi-major axis of 300~au). Planet masses range from \num{$0.012~\, M_{\rm Jup}$\ to $7~\, M_{\rm Jup}$} with a median planet mass of \num{$0.4\, M_{\rm Jup}$}.  

We also show the cumulative distributions in spectral type bins (M/K/G/F/Massive) according to the ranges given in Table \ref{tab:st_mass} and the minimum and maximum host masses in the sample. The host stars in \citet{Pearce22} have masses ranging from $0.22\, M_{\odot}$ to $3.7\, M_{\odot}$ but are heavily biased toward stars with masses exceeding $0.8\, M_{\odot}$; the sample contains only four M dwarfs. The inferred planet minimum mass distribution is a strong function of host mass. However, after shifting all the distributions by the median planet mass (bottom panels),  a Kolmogorov-Smirnov test shows no significant differences between them.

The top panel of Figure \ref{fig:pearce} compares various power-laws to the \citet{Pearce22} cumulative distribution. The power-laws are normalized to match the \citet{Pearce22} distribution at \num{$\log (m_{\rm p} / M_{\rm Jup}) = -0.7~\mathrm{and}~ 0.0$ ($0.2~M_{\rm Jup}~\mathrm{and}~1~M_{\rm Jup}$)}.  A power-law index of $p=1.0$, as suggested by the $\mu$FFP population, is clearly too steep for the inferred debris disk population. This difference could be resolved if there is a large population of small objects found as $\mu$FFPs that are not associated with a debris disk structure.

Table \ref{tab:pearce_spt} lists planet frequencies for the \citet{Pearce22} population. We count the number of planets ($N_{\rm pl}$) in a given planet mass ($m_p$) and spectral type (SpT) bin and calculate the fraction of planets (frac) of this type. We weight this frequency by the fraction of the stellar population ``Pop Frac." represented by the range of host masses in that bin ($min ~ M_*, max ~ M_*$) to derive the predicted number of planets per 100 stars (``All"). This estimate would be the total planet frequency if we assume the sample of stars with debris disks is representative of the planet population around all stars. We also consider an additional weight to account for the debris disk frequency, assuming DD Frac. = \num{10\%, 25\%, and 50\%} for M, FGK, and AB stars, respectively \citep{Lestrade09}. The resulting values are given in the ``DD-only" column and are the minimum frequencies assuming that stars without debris disks have no planets.
We sum the values to obtain the total predicted number of planets in various mass bins, which are given in Table \ref{tab:pearce_all}.

For the mass ranges $1~M_{\rm Jup} < m_{\rm p} < 13~M_{\rm Jup}$ and $1~M_{\rm Sat} < m_{\rm p} < 1~M_{\rm Jup}$, the frequency is \num{3.5 and 8.1 planets per 100 stars}, respectively, for the \citet{Pearce22} sample. These values are similar to the values from the $\mu$FFP populations (see Table \ref{tab:ffp_freq}). However, they require that all of the $\mu$FFPs in this range are bound planets, although this requirement can be relaxed if only stars with debris disks have planets of this type.

The \citet{Pearce22} masses are lower limits on the masses of the planets. Because the estimates of the \citet{Pearce22} planet frequency for $m_{\rm p} > M_{\rm Sat}$ are already roughly equivalent to the $\mu$FFP population, it is difficult for them to be significantly more massive than the minimum values. As an extreme example, we predict \num{16.2} total planets with $1~M_{\rm Earth} < m_{\rm p} < 13~M_{\rm Jup}$ per 100 stars from \citet{Pearce22}. If all of them actually had masses between $1~M_{\rm Jup}$ and $13~M_{\rm Jup}$, that would vastly exceed the allowable totals from the $\mu$FFP population. We estimate that \num{$\lesssim 25\%$} of the \citet{Pearce22} planets with minimum masses $< 1~M_{\rm Jup}$ can have true masses $> 1~M_{\rm Jup}$, even including the assumption that all planets in the $1~M_{\rm Jup} < m_{\rm p} < 13~M_{\rm Jup}$ bin have true masses $>13~M_{\rm Jup}$. Thus, the $\mu$FFP population can place upper limits on the masses of the planets predicted from debris disk physics.

The alternative is that the inner edges of the disks are sculpted by multiple planets. \citet{Pearce22} conclude that the minimum masses for multiple, equal mass planets to maintain the inner edge of the disk are generally at least an order of magnitude smaller than for a single planet (see Figure \ref{fig:pearce_Mpn}), and the majority of the planets have masses $m_{\rm p} < 1 M_{\rm Sat}$. \citet{Pearce22} do not calculate the total number of planets required for each disk, so we cannot directly compare the frequencies or the resulting cumulative distributions with the microlensing results. Depending on how many planets are required, this scenario may be more compatible with the microlensing results.

\vskip 4ex
{\subsection{ALMA Substructure}\label{sec:alma}}

\begin{deluxetable}{rr|rr|rr|rr|rr|rr|}
    \tablecaption{Planets per 100 Stars Inferred from ALMA Disks\label{tab:alma}}
    \tabletypesize{\footnotesize}
    \tablehead{}
    \startdata
\multicolumn{12}{c}{\citet{Zhang18}}\\
\hline
\multicolumn{2}{c|}{}              & \multicolumn{2}{c|}{$\alpha=10^{-2}$} & \multicolumn{2}{c|}{$\alpha=10^{-3}$} & \multicolumn{2}{c|}{$\alpha=10^{-4}$} \\
\hline
\multicolumn{2}{l|}{Total Planets:} & \multicolumn{2}{c|}{11} & \multicolumn{2}{c|}{15} & \multicolumn{2}{c|}{15} \\
\multicolumn{2}{l|}{Total Stars:}   & \multicolumn{2}{c|}{14} & \multicolumn{2}{c|}{14} & \multicolumn{2}{c|}{14} \\
\hline
\multicolumn{2}{l|}{Mass Range}                        & $N_{\rm pl}$ &   freq  & $N_{\rm pl}$ &   freq  & $N_{\rm pl}$ &   freq  \\
\hline
\multicolumn{2}{l|}{$[1\,M_{\rm Jup}, 13\,M_{\rm Jup}]$} &            5 &   35.7  &            2 &   14.3  &            3 &   21.4  \\
\multicolumn{2}{l|}{$[1\,M_{\rm Sat},  1\,M_{\rm Jup}]$} &            2 &   14.3  &            5 &   35.7  &            3 &   21.4  \\
\multicolumn{2}{l|}{$[1\,M_{\rm Nep},  1\,M_{\rm Sat}]$} &            3 &   21.4  &            4 &   28.6  &            5 &   35.7  \\
\multicolumn{2}{l|}{$[1\,M_{\rm Earth}, 1\,M_{\rm Nep}]$} &            0 &    0.0  &            3 &   21.4  &            4 &   28.6  \\
\hline
\multicolumn{2}{l|}{$[5\,M_{\rm Jup}, 13\,M_{\rm Jup}]$} &            0 &    0.0  &            0 &    0.0  &            1 &    7.1  \\
\hline
\hline
\multicolumn{12}{c}{\citet{Lodato19}}\\
\hline
\multicolumn{2}{c|}{}              & \multicolumn{2}{c|}{T1: Taurus} & \multicolumn{2}{c|}{T2: \citet{Zhang18}} & \multicolumn{2}{c|}{T3: \citet{Bae18}} & \multicolumn{2}{c|}{Combined} \\
\hline
\multicolumn{2}{l|}{Total Planets:} & \multicolumn{2}{c|}{15} & \multicolumn{2}{c|}{18} & \multicolumn{2}{c|}{13} & \multicolumn{2}{c|}{46} \\
\multicolumn{2}{l|}{Total Stars:}   & \multicolumn{2}{c|}{10} & \multicolumn{2}{c|}{14} & \multicolumn{2}{c|}{ 9} & \multicolumn{2}{c|}{33} \\
\hline
\multicolumn{2}{l|}{Mass Range}                        & $N_{\rm pl}$ &   freq  & $N_{\rm pl}$ &   freq  & $N_{\rm pl}$ &   freq  & $N_{\rm pl}$ &   freq  \\
\hline
\multicolumn{2}{l|}{$[1\,M_{\rm Jup}, 13\,M_{\rm Jup}]$} &            2 &   20.0  &            2 &   14.3  &            2 &   22.2  &            6 &   18.2  \\
\multicolumn{2}{l|}{$[1\,M_{\rm Sat},  1\,M_{\rm Jup}]$} &            3 &   30.0  &            5 &   35.7  &            4 &   44.4  &           12 &   36.4  \\
\multicolumn{2}{l|}{$[1\,M_{\rm Nep},  1\,M_{\rm Sat}]$} &            7 &   70.0  &            3 &   21.4  &            6 &   66.7  &           16 &   48.5  \\
\multicolumn{2}{l|}{$[1\,M_{\rm Earth}, 1\,M_{\rm Nep}]$} &            2 &   20.0  &            7 &   50.0  &            0 &    0.0  &            9 &   27.3  \\
\hline
\multicolumn{2}{l|}{$[5\,M_{\rm Jup}, 13\,M_{\rm Jup}]$} &            1 &   10.0  &            0 &    0.0  &            0 &    0.0  &            1 &    3.0  \\
\hline
\hline
\multicolumn{12}{c}{\citet{Wang21}\tablenotemark{a}}\\
\hline
\multicolumn{2}{c|}{}              & \multicolumn{2}{c|}{$\alpha=10^{-2}$, dust} & \multicolumn{2}{c|}{$\alpha=10^{-2}$, gas} & \multicolumn{2}{c|}{$\alpha=10^{-3}$, dust} & \multicolumn{2}{c|}{$\alpha=10^{-3}$, gas} & \multicolumn{2}{c|}{$\alpha=10^{-4}$} \\
\hline
\multicolumn{2}{l|}{Total Planets:} & \multicolumn{2}{c|}{54} & \multicolumn{2}{c|}{54} & \multicolumn{2}{c|}{54} & \multicolumn{2}{c|}{54} & \multicolumn{2}{c|}{54} \\
\multicolumn{2}{l|}{Total Stars:}   & \multicolumn{2}{c|}{35} & \multicolumn{2}{c|}{35} & \multicolumn{2}{c|}{35} & \multicolumn{2}{c|}{35} & \multicolumn{2}{c|}{35} \\
\hline
\multicolumn{2}{l|}{Mass Range}                        & $N_{\rm pl}$ &   freq  & $N_{\rm pl}$ &   freq  & $N_{\rm pl}$ &   freq  & $N_{\rm pl}$ &   freq  & $N_{\rm pl}$ &   freq  \\
\hline
\multicolumn{2}{l|}{$[1\,M_{\rm Jup}, 13\,M_{\rm Jup}]$} &            9 &   25.7  &           14 &   40.0  &            6 &   17.1  &            6 &   17.1  &            1 &    2.9  \\
\multicolumn{2}{l|}{$[1\,M_{\rm Sat},  1\,M_{\rm Jup}]$} &            9 &   25.7  &           15 &   42.9  &            9 &   25.7  &            9 &   25.7  &            7 &   20.0  \\
\multicolumn{2}{l|}{$[1\,M_{\rm Nep},  1\,M_{\rm Sat}]$} &           16 &   45.7  &            8 &   22.9  &           20 &   57.1  &           20 &   57.1  &           26 &   74.3  \\
\multicolumn{2}{l|}{$[1\,M_{\rm Earth}, 1\,M_{\rm Nep}]$} &           19 &   54.3  &           16 &   45.7  &           19 &   54.3  &           19 &   54.3  &           20 &   57.1  \\
\hline
\multicolumn{2}{l|}{$[5\,M_{\rm Jup}, 13\,M_{\rm Jup}]$} &            2 &    5.7  &            2 &    5.7  &            0 &    0.0  &            0 &    0.0  &            0 &    0.0  \\
\hline
\hline
\multicolumn{12}{c}{\citet{Zhang23}}\\
\hline
\multicolumn{2}{c|}{}              & \multicolumn{2}{c|}{$\alpha_{\rm max} = 0.1~\mathrm{mm}$} & \multicolumn{2}{c|}{$\alpha_{\rm max} = 1~\mathrm{mm}$} & \multicolumn{2}{c|}{$\alpha_{\rm max} = 1~\mathrm{cm}$} \\
\hline
\multicolumn{2}{l|}{Total Planets:} & \multicolumn{2}{c|}{24} & \multicolumn{2}{c|}{22} & \multicolumn{2}{c|}{10} \\
\multicolumn{2}{l|}{Total Stars:}   & \multicolumn{2}{c|}{15} & \multicolumn{2}{c|}{15} & \multicolumn{2}{c|}{15} \\
\hline
\multicolumn{2}{l|}{Mass Range}                        & $N_{\rm pl}$ &   freq  & $N_{\rm pl}$ &   freq  & $N_{\rm pl}$ &   freq  \\
\hline
\multicolumn{2}{l|}{$[1\,M_{\rm Jup}, 13\,M_{\rm Jup}]$} &            2 &   13.3  &            2 &   13.3  &            0 &    0.0  \\
\multicolumn{2}{l|}{$[1\,M_{\rm Sat},  1\,M_{\rm Jup}]$} &            3 &   20.0  &            2 &   13.3  &            1 &    6.7  \\
\multicolumn{2}{l|}{$[1\,M_{\rm Nep},  1\,M_{\rm Sat}]$} &            8 &   53.3  &            8 &   53.3  &            4 &   26.7  \\
\multicolumn{2}{l|}{$[1\,M_{\rm Earth}, 1\,M_{\rm Nep}]$} &            8 &   53.3  &            9 &   60.0  &            4 &   26.7  \\
\hline
\multicolumn{2}{l|}{$[5\,M_{\rm Jup}, 13\,M_{\rm Jup}]$} &            0 &    0.0  &            0 &    0.0  &            0 &    0.0  \\
\hline
\hline
\enddata
\tablenotetext{a}{In this table, the choice of ``dust" vs. ``gas" refers only to ambiguous gaps, which may be in either category. See text.}
\tablecomments{Because some of the inferred companions have $m_{\rm p} > 
13~M_{\rm Jup}$, ``Total Planets" may exceed the sum of the $N_{\rm pl}$ column.}
\end{deluxetable}

\begin{figure}
    \includegraphics[width=0.95\textwidth]{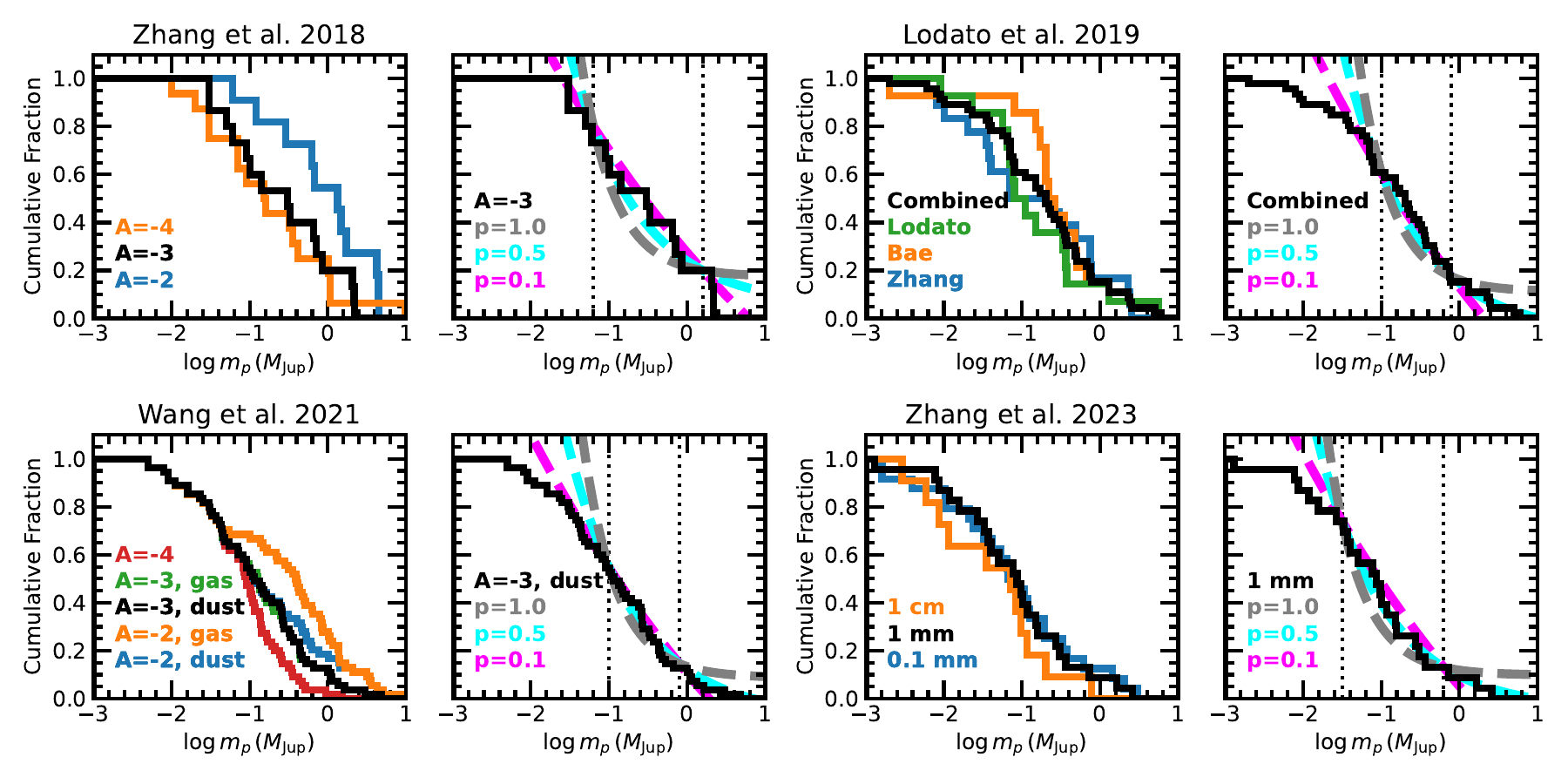}
    \caption{Cumulative distribution functions of planets inferred from ALMA disks from \citet{Zhang18} (upper left), \citet{Lodato19} (upper right), \citet{Wang21} (lower left), and \citet{Zhang23} (lower right). The left panel in each pair shows distributions for different assumptions of physics from each study (for \citealt{Zhang18} and \citealt{Wang21}: $A\equiv \log\alpha$, \citealt{Zhang23}: different limits for the maximum grain size) or, in the case of \citet{Lodato19}, different observational samples (see legend). The right panel shows one of those distributions relative to power-laws with different indices $p$, normalized to the points indicated by the vertical dotted lines. \label{fig:alma_dist}}
\end{figure}

As summarized in \citet{Najita22}, recent ALMA observations divide
protostellar disks around solar-type stars into two general classes
\citep[see also][and references therein]{Alma15,Avenhaus18,Huang18a,
Long19,Cieza19,vanderMarel2021b,Bae23_PPVII}.
The youngest solar-type stars all have a compact disk with a radius of
20--30~au.  Roughly 25\% of these young stars have thin rings of gas and dust
at larger radii.  Rings have typical widths, $\delta a /a \approx$ 0.05--0.10, 
and lie at distances $a \approx$ 40--200~au from the host star
\citep[and references therein]{Hughes18,Bae23_PPVII,Michel2023,Huang2024}.  For robust ALMA
detections, the rings are massive, $\gtrsim$ 5--$10\,M_{\oplus}$ of solids in
the form of mm-sized to cm-sized pebbles.
There are also rings in transitional disk systems, where the compact disk
has an inner hole \citep[e.g.,][and references therein]{Bae23_PPVII}.
A small sample of protostellar disks have spirals or crescents \citep{vanderMarel2021a,Bae23_PPVII}.

The gaps seen in ALMA disks probe the planet population at a much younger age \citep[see also][]{Andrews2016,Sheehan2020,Teague2021}. In this section, we consider \num{four} works that calculate properties for planets required to open the observed gaps in protoplanetary disks: \citet{Zhang18} (Section \ref{sec:zhang18}), \citet{Lodato19} (Section \ref{sec:lodato19}), \citet{Wang21} (Section \ref{sec:wang21}), and \citet{Zhang23} (Section \ref{sec:zhang23}). We discuss our conclusions in Section \ref{sec:alma_conclusions}.

For the \num{four} studies discussed below, we calculate the expected number of planets per 100 stars for each study. Each sample spans a range of stellar masses. However, the total number of stars in each sample is small, and there does not appear to be a strong correlation between inferred planet mass and host mass. Hence, we do not re-weight the sample by spectral type/stellar mass bin, but instead assume the frequency is independent of stellar mass. In addition, for consistency with previous calculations, we limit the proposed planets to those in gaps at \num{$R_{\rm gap} \ge 10~\mathrm{au}$}. The calculations of the planet frequencies are given in Table \ref{tab:alma} and the cumulative distributions of the inferred planet populations are shown in Figure \ref{fig:alma_dist}. 

\vspace{12pt}
{\subsubsection{Zhang et al (2018)}
\label{sec:zhang18}}

\citet{Zhang18} took a sample of 14 protoplanetary disks from the DSHARP survey \citep{Andrews18} containing a total of 19 gaps. Using hydrodynamical simulations under different assumptions for the disk turbulent viscosity coefficient, $\alpha$, they map the properties of the gaps to planet masses (for the legend in Figure \ref{fig:alma_dist}, we use $A \equiv \log \alpha$.). In this sample, there are \num{40--50 planets per 100 stars} with $a \ge 10~$ au between $1~M_{\rm Sat}$ and $13~M_{\rm Jup}$  regardless of the assumed value of $\alpha$. This frequency is well in excess of the number allowed by the $\mu$FFP population in this mass range: \num{7.5 \citep{Gould22_FFPs} and 11.7 \citep{Sumi23_FFPs} planets per 100 stars}.

\vspace{12pt}
{\subsubsection{Lodato et al (2019)}
\label{sec:lodato19}}

Instead of using detailed hydrodynamical simulations, \citet{Lodato19} use a simple assumption that the gap width scales with the Hill radius of a planet. They calculate the masses of the hypothetical planets existing in several, non-overlapping, samples of observed proto-planetary disks: their own sample of disks from Taurus, the DSHARP sample from \citet{Zhang18}, and a compilation of gaps from \citet{Bae18}. Table \ref{tab:alma}, lists the number of planets per star for each of their samples and also for the combination. As with the \citet{Zhang18} sample, the number of planets per star between $1~M_{\rm Sat}$ and $13~M_{\rm Jup}$ is well in excess of the number allowed by the $\mu$FFP populations.

\vspace{12pt}
{\subsubsection{Wang et al (2021)}
\label{sec:wang21}}

\citet{Wang21} take the calculations of \citet{Lodato19} one step further and consider whether the observed gap in the dust disk is likely to have a corresponding gap in the gas disk. Planets below a certain mass only open gaps in the dust disk; the Hill radius relation of \citet{Lodato19} should apply (the ``dust" regime). However, planets above this mass are likely to open gaps in the gas disk and are therefore governed by the empirical relation of \citet{Kanagawa16} (the ``gas" regime). That relation generally implies more massive planets than the Hill radius relation for the same gap width. 

For each dust gap in their sample, \citet{Wang21} classify it according to which regime applies or whether it is ambiguous. For our calculations of the number of planets per star, we perform two sets of calculations for each value of $\alpha$. First, we assume that the ambiguous gaps are all in the ``dust" regime, and then we assume that they are all in the ``gas" regime. Thus, the labels in Figure \ref{fig:alma_dist} and Table \ref{tab:alma} refer to the distinction only for the gaps in the ambiguous regime, rather than completely separating the populations by the assumed underlying physics. For the $A \equiv\log\alpha = -4$ regime, there are no ambiguous gaps, so we give only one set of values.

Regardless of the choice for ambiguous planets, there is significant tension with the $\mu$FFP population. The microlensing results allow for \num{2.4 \citep{Gould22_FFPs} and 3.4 \citet{Sumi23_FFPs}} large gas giants ($1~M_{\rm Jup} < m_{\rm p} < 13~M_{\rm Jup}$) and \num{5.1 and 8.2} medium
gas giants ($1~M_{\rm Sat}< m_{\rm p} <~M_{\rm Jup}$) per 100 stars, respectively. Even with Poisson statistics and regardless of the assumed physics, the inferred gap populations from \citet{Wang21} imply \num{too many} large planets relative to the $\mu$FFP population.

\vspace{12pt}
{\subsubsection{Zhang et al (2023)}
\label{sec:zhang23}}

\citet{Zhang23} calculate the inferred mass of planets for gaps in disks in Taurus using the framework of \citet{Zhang18}. In our calculations of the planet frequencies, in contrast to \citet{Zhang23}, we do not exclude any of the proposed planets from the sample, so our numbers may differ from those reported in the original paper. For this sample, the results are consistent with the microlensing FFP population, but only under the assumption that the maximum grain size is $\alpha_{\rm max} = 1~$ cm. Smaller $\alpha_{\rm max}$ leads to larger planets and strong tension with the $\mu$FFP population. 

\vspace{12pt}
{\subsubsection{ALMA Conclusions}
\label{sec:alma_conclusions}}

The populations of planets inferred from ALMA gaps prefer a much shallower power-law index ($p \sim 0.5$) than the $\mu$FFP population (Figure \ref{fig:alma_dist}). There are also significantly more giant planets predicted by ALMA gaps than are compatible with the $\mu$FFP population. Unlike for direct imaging or radial velocity planets, because these inferred planets are embedded in a protoplanetary disk, they cannot be explained by assigning them to the stellar mass function.

There are several ways this tension might be resolved. First, we have assumed that the inferred ALMA planet populations are representative of the planet population as a whole. \citealt{Lodato19} argues the most significant bias in the Taurus sample is against massive disks (which are likely to have additional massive planets), which supports that assumption. On the other hand, they argue that only 35\% of young stars have disks with gaps. This factor of $\sim 3$ would reduce, but not entirely resolve, the tension with the microlensing results for certain physical assumptions. However, it also requires that planets with $a > 10~$ au and $m_p > M_{\rm Sat}$ are {\em only} found around stars with disks with gaps. 

Another explanation could be that as the systems evolve, the planets migrate away from where they are inferred to form in the protoplanetary disk. However, the migration cannot be outward because those planets would still be detected as part of the $\mu$FFP population, even if they are ejected from their host systems. Inward migration would likely require that a significant fraction are absorbed by the star because the number of giant planets at $a < 10~\mathrm{au}$ has been measured by multiple studies to be $\ll 1$ per star \citep{Fernandes19, Suzuki16}.

Finally, this tension could suggest problems with the assumed physics of gap formation by planets in these studies. Perhaps there is not a 1-to-1 correspondence between planets and gaps. For example, a single planet may explain multiple gaps in the disk around AS 209 \citep{Fedele18_AS209}.
Alternatively, theories about the physical mechanisms controlling gap formation could be incomplete. Regardless, the tension indicates that the $\mu$FFP population has a significant ability to constrain these theories and assumptions.

\section{Conclusions}

Data from microlensing \citep{Suzuki16, Poleski21}, radial velocities \citep{Cumming08, Mayor11, Bonfils13}, and transits \citep{Lissauer2023} demonstrate that planets with masses comparable to Saturn and Jupiter are significantly less common than planets with masses comparable to Earth and Neptune. The declining frequency of planets with increasing mass mirrors a similar phenomenon among stars \citep[e.g.,][]{Salpeter1955,MillerScalo1979,Chabrier05,Kroupa2013,Kirkpatrick2024}. From the most massive O-type stars to the lowest mass brown dwarfs, the frequency is a declining function of mass. For stars more massive than $\sim$ 0.1--0.2~$M_\odot$, all derivations of the initial mass function (IMF) agree reasonably well. However, among the lowest mass hydrogen-burning stars and lower mass brown dwarfs, different formulations of the IMF diverge. There is an open question of whether or not the stellar IMF continues into the brown dwarf and planetary mass regime, e.g., as suggested by the discovery of isolated planetary-mass cores in young star-forming regions \citep[e.g.,][]{Pearson21_NGC2264, Damian23_sigmaOri}.

The mass function (MF) of isolated objects identified with microlensing can be decomposed into two separate components: a free-floating planet (FFP) component and a stellar component \citep{Gould22_FFPs, Sumi23_FFPs}. We show that the stellar MF from \citet{Sumi23_FFPs} has significantly more objects with $M < 0.1~M_\odot$ than the \citet{Chabrier05} MF, but is similar to the local IMF derived by \citet{Kirkpatrick2024} (see Figure \ref{fig:mfs}). The \citet{Sumi23_FFPs} MFs imply that objects with $M \gtrsim 1~M_{\rm Jup}$ are dominated by objects from the low-mass tail of the stellar mass function, with relatively few objects coming from the high-mass end of planet formation. The conclusions are similar using the \citet{Chabrier05} stellar MF and the \citet{Gould22_FFPs} microlensing FFP ($\mu$FFP) MF, but the transition point shifts to $M \gtrsim 6~M_{\rm Jup}$.

We compared the microlensing MFs with the populations of companions with masses $m_{\rm p} > 1~M_{\rm Jup}$ and semi-major axes $a > 10~\mathrm{au}$ observed by direct imaging \citep[][Section \ref{sec:di}]{Nielsen19, Vigan21} and radial velocity \citep[][Section \ref{sec:rv}]{Fernandes19}. The frequency of these companions is equal to the frequency in the $\mu$FFP population. This result implies that all $\mu$FFPs with these masses are bound planets for which there is no measurable microlensing signal from the host star as hypothesized by \citet{Gould22_FFPs}. These planets can only exist around more massive stars; there cannot be additional undetected companions around stars outside the ranges probed by those studies. Finally, the microlensing objects from the stellar MF in this mass range should not have host stars, unless the hosts would be undetectable with existing radial velocity or direct imaging surveys. 

We predict that if high-resolution imaging is taken for $\mu$FFP candidates with $M > 1~M_{\rm Jup}$, there must be at least as many host stars with masses $\gtrsim 1~M_{\odot}$ as the number of such candidates predicted to belong to the planetary, rather than stellar, component of the microlensing MF. Alternatively, some fraction of the direct imaging and radial velocity companions are actually brown dwarfs, which formed through the star formation process.

We also conclude that the number of planets with minimum masses $>1~M_{\rm Sat}$ inferred from debris disks assuming a single planet maintains the inner edge of the disk \citep[][Section \ref{sec:disks}]{Pearce22}  is roughly equal to the number of such planets in the $\mu$FFP MF. This result agrees with our analysis comparing the $\mu$FFP population with radial velocity and direct imaging populations but extending to lower masses: either all $\mu$FFPs must be bound or stars without debris disks tend not to host planets with masses $>1 ~M_{\rm Sat}$ and orbits with $a > 10~\mathrm{au}$. 

The inferred debris disk planet population extends below $1~M_{\rm Sat}$. The power-law index for these planets has a much shallower slope than the $\mu$FFP MF. The $\mu$FFP population can easily accommodate this population of smaller planets and additional contributions from smaller planets, including those around stars without debris disks or ejected planets. At the same time, the planet masses from \citet{Pearce22} are lower limits, but if they were much larger, they would violate the constraints from the higher mass $\mu$FFP population. Hence, the $\mu$FFP population constrains the masses of the population of planets sculpting debris disks or suggests that they are scuplted by multiple, smaller planets, an alternative hypothesis investigated by \citet{Pearce22}.

In addition, the $\mu$FFP population is not consistent with the hypothesis that every ALMA gap contains a planet \citep[][Section \ref{sec:alma}]{Zhang18,Lodato19,Wang21,Zhang23}. Even if we account for the possibility that only stars with massive disks host such planets, the required numbers of $m_{\rm p} > 1 ~M_{\rm Sat}$ are in tension with the $\mu$FFP population. Because the $\mu$FFP population constrains all objects with $a>10~\mathrm{au}$, migration can only resolve this tension if the planets migrate inward. In that case, a large fraction would have to be absorbed by their stars due to the constraints on the frequency of giant planets with $a < 10~\mathrm{au}$. More plausibly, there is not a 1-to-1 correspondence between ALMA gaps and planets. Instead, these results favor a scenario in which a single planet induces multiple gaps.

These comparisons demonstrate the power of microlensing measurements of the mass function to constrain the wide-orbit planet population and the physics governing structures in circumstellar disks. Stronger comparisons would be possible with larger samples of direct imaging and radial velocity planets, which would allow more precise frequency measurements as a function of companion mass, and a better understanding of the range of host star masses that are probed by direct imaging and ALMA disk studies.

\section*{Acknowledgements}
J.C.Y. acknowledges support from U.S. NSF Grant No. AST-2108414. 

This research has made use of the NASA Exoplanet Archive, which is operated by the California Institute of Technology, under contract with the National Aeronautics and Space Administration under the Exoplanet Exploration Program.

\bibliography{draft_v5.bbl}
\end{document}